\documentclass[twocolumn,showpacs,preprintnumbers,amsmath,amssymb, prb]{revtex4-1}

\usepackage{stmaryrd}
\usepackage{txfonts}
\usepackage{amssymb}
\usepackage{mathrsfs}
\usepackage{graphicx}
\usepackage{dcolumn}
\usepackage{bm}
\usepackage{epsfig}
\usepackage{color}                    
\usepackage{hyperref}                 

\begin{document}
\preprint{\href{http://dx.doi.org/10.1103/PhysRevB.86.224513}{L. N. Bulaevskii and S. Z. Lin, Phys. Rev. {\bf{B}} {\bf 86}, 224513 (2012).}}

\title{Self-induced pinning of vortices in the presence of ac driving force in magnetic superconductors}

\author{Lev N. Bulaevskii and Shi-Zeng Lin}
\affiliation{Theoretical Division, Los Alamos National Laboratory, Los Alamos, New Mexico 87545}

\begin{abstract}
 We derive the response of the magnetic superconductors in the vortex state to the ac Lorentz force, $F_L(t)=F_{{\rm ac}}\sin(\omega t)$, taking into account the
interaction of vortices with the magnetic moments described by the relaxation dynamics (polaronic effect). At low amplitudes of the driving force  $F_{{\rm ac}}$ the dissipation in the system is suppressed due to the enhancement of the effective viscosity at low frequencies and due to formation of the magnetic pinning at high frequencies $\omega$.  In the adiabatic limit with low frequencies $\omega$ and high amplitude of the driving force $F_{ac}$, the vortex and magnetic polarization form a vortex polaron when $F_L(t)$ is small. When $F_L$ increases, the vortex polaron accelerates and at a threshold driving force,  the vortex polaron dissociates and the motion of vortex and the relaxation of magnetization are decoupled. When $F_L$ decreases, the vortex is retrapped by the background of remnant magnetization and they again form vortex polaron. This process repeats when $F_L(t)$ increases in the opposite direction. Remarkably, after dissociation, decoupled vortices move in the periodic potential induced by magnetization which remains for some  periods of time due to retardation after the decoupling. At this stage vortices oscillate with high frequencies determined by the Lorentz force at the moment of dissociation.  We derive also the creep rate of vortices and show that magnetic moments suppress creep rate. 
 \end{abstract}
\date{\today}
\maketitle

\section{Introduction}
To explain behavior of the critical current in Er borocarbide $\rm{ErNi_2B_2C}$ \cite{Canfield98,Budko06,Gupta2006,Gammel2000,James2001} it was proposed in Ref.~\onlinecite{Bulaevskii12a} that in magnetic superconductors vortex motion becomes suppressed due to their interaction 
with the nonuniform part of magnetic moment polarization. Magnetic field in  the vortex state is nonuniform and so is the  polarization of magnetic moments. 
In the flux flow state, vortices, to move, need to change polarization as they are attracted to the polarization clouds induced at their previous positions. Due to nonzero magnetic relaxation time vortex viscosity is enhanced and this enhancement increases with magnetic relaxation time.
The vortex together with accompanied nonuniform polarization form a vortex polaron at low vortex velocity. As bias current and Lorentz force increase so does vortex velocity and above some critical velocity vortices get rid of polarization and move at higher velocities defined by the usual Bardeen-Stephen drag coefficient. 
At this polaron dissociation voltage jumps to a higher value and from the point of view of I-V characteristics the corresponding bias current may be identified as the critical current. When pinning due to defects is present, vortices start to move with enhanced viscosity and low dissipation above the critical current, determined by the pinning due to defects. Then above the dissociation critical current they move with the velocity determined by the Bardeen-Stephen drag coefficient \cite{Lin12m}. Thus effectively the critical current is the sum of standard pinning critical current and polaron dissociation critical current.  The
polaronic mechanism constraining vortex motion is effective not only in crystals, but in multilayered systems of superconductor and magnetic layers whose thickness is smaller or of the order of the superconducting penetration depth $\lambda$; see Ref.~\onlinecite{Lin12m}.
 
The polaronic mechanism is inherent to any magnetic superconductor, but the strength of magnetic coupling of vortex with magnetic moments depends
crucially on the London penetration length which determines the nonuniform part of the magnetic field inside superconductors. The interaction 
drops as $1/\lambda^4$ as $\lambda$ increases and thus the polaronic mechanism is most effective in superconductors with low $\lambda$'s 
like borocarbides where $\lambda$ is of the order 500 $\AA$. Another important parameter is the magnetic susceptibility $\chi=dM/dH$ ($M$ is the magnetization) at the 
operating applied magnetic field $H$. Finally the third crucial parameter in slowing down vortex motion is the relaxation time $\tau$ of the magnetic moments.

In this paper we consider vortex motion in the presence of the alternating in time Lorentz force and show that the polaronic effect results 
in suppression of vortex oscillations and, consequently, dissipation. We consider also thermally activated vortex motion via the potential barrier inherent to vortex creep at low temperatures in the Andersen-Kim model. We show that creep rate is also suppressed by the polaronic effect
resulting in lower dissipation at low temperatures. We predict that a strong ac Lorentz force with frequency $\omega\sim 1/(\chi\tau)$ results in vortex lattice oscillations with the frequency $\Omega\gg\omega$ just after polaron dissociation
due to retardation in response of magnetic moments to the inhomogeneous magnetic field of moving vortex lattice. 

\section{General equations}

We consider the system of vortices at the coordinates ${\bf R}_i=(x_i, y_i)$  interacting with the magnetic moments described by the magnetization $M({\bf r})$ in thin film with the thickness $d$ much smaller than the London penetration length $\lambda$. The Lagrangian ${\cal L}\{{\bf R}_i(t), M({\bf r},t)\}$  for the whole system reads 
\begin{align}\label{eq1}
\nonumber{\cal L}\{{\bf R}_i(t), M({\bf r},t)\}=
{\cal L}_M\{M({\bf r},t)\}+
{\cal L}_v\{{\bf R}_i(t)\}\\+{\cal L}_{{\rm int}}\{M({\bf r},t),{\bf R}_i(t)\}+{\cal L}_{vv}\{\mathbf{R}_i\}+\mathcal{L}_F\{\mathbf{J}\}. 
\end{align}
The Lagrangian for the magnetic subsystem is
\begin{align}\label{eq2}
{\cal L}_M\{M({\bf r},t)\}=-d\int d{\bf r}M^2({\bf r},t)/(2\chi).
\end{align}
The Lagrangian for the interaction between vortices and pinning potential due to quenched disorder reads
\begin{align}\label{eq3}
{\cal L}_v\{{\bf R}_i(t), {\bf r}_j\}=-\sum_{i,j}U({\bf R}_i-{\bf r}_j),
\end{align}
where $U({\bf R}_i-\mathbf{r}_j)$ is the pinning potential at ${\bf r}_j$.  Further, ${\cal L}_{vv}\{\mathbf{R}_i\}$ is the vortex-vortex interaction and $\mathcal{L}_F\{\mathbf{J}\}=\sum_i\mathbf{J}\cdot{\mathbf{R}_i}\Phi_0d/c$ is the Lagrangian due to Lorentz force in the presence of bias current density $\mathbf{J}$, while $\Phi_0=h c/(2e)$ is the flux quantum. Here $\chi$ is the magnetic susceptibility at the working external magnetic field. It describes response of magnetic moments to nonuniform component of the field induced by vortices. We describe the magnetic moments in the continuous approximation via the magnetization $M({\mathbf r},t)$, because distance between spins is much smaller than the superconducting penetration length $\lambda$.  We consider the Ising spins with the moments along the vortex direction, as in Er borocarbide. We ignore the pair breaking effect of the magnetic moments. The interaction between vortices at the coordinates ${\mathbf R}_i$, and the  magnetic moments is determined by the term 
\begin{equation}\label{eq4}
{\cal L}_{\rm{int}}\{{\mathbf R}_i,{\mathbf M}\}=d\int d{\mathbf r} B(\mathbf{R}_i-\mathbf {r},t) M({\mathbf r},t).
\end{equation}
Both the vortices and magnetization are described by a relaxation dynamics characterized by the dissipation function ${\cal R}\{{\bf R}_i(t), M({\bf r},t)\}={\cal R}_M+{\cal R}_v$, where  
\begin{align}\label{eq5}
{\cal R}_M\{\dot{M}({\bf r})\}=\frac{1}{2}\tau d\int d{\bf r}\dot{M}^2({\bf r}), 
\ \ \ \ {\cal R}_v\{\dot{{\bf R}_i}\}=\eta d\sum_i\frac{1}{2}\dot{{\bf R}}_i^2.
\end{align}
Here $\tau$ is the relaxation time for a single spin and $\eta$ is the Bardeen-Stephen drag coefficient per unit vortex length,  $\eta={\Phi _0^2}/({2\pi  \xi ^2c^2\rho_n })$ with $\rho_n$ the normal resistivity slightly above $T_c$. Within the London approximation, the magnetic field $B({\bf r})$ induced by vortices is given by $B({\bf r})=1/(2\pi)^2\sum_i\int d{\bf k}B({\bf k}\exp[i{\bf k}({\bf r}- \mathbf{R_i})]$, ${\bf k}=(k_x,k_y)$ with
\begin{align}\label{eq6}
 B({\bf k})=\frac{\Phi_0}{1+\lambda^2{\bf k}^2}.
\end{align}
Note, that the superconducting penetration length here is the renormalized length given by the expression $\lambda^2=\lambda_L^2(1-4\pi\chi)$, where $\lambda_L$ describes magnetic field penetration in the absence of the magnetic moments.\cite{Tachiki79,Gray83,Buzdin84,Bulaevskii85} The equation of motion for vortices is
\begin{equation}\label{eq7}
\frac{d}{dt}\frac{\delta {\cal L}}{\delta {\dot{\bf R}}_i}-\frac{\delta {\cal L}}{\delta {\bf R}_i}+\frac{\delta {\cal R}}{\delta \dot{{\bf R}}_i}=0.
\end{equation}
It gives
\begin{align}\label{eq8}
\nonumber\eta d\frac{\partial {\mathbf R}_i}{\partial t}=\frac{\partial{\cal L}_{\rm{vv}}\{{\mathbf R}_i,{\mathbf R}_j\}}{\partial \mathbf{R}_i}+\frac{\partial {\cal L}_{\rm{int}}\{{\mathbf R}_i,{\mathbf M}\}}{\partial {\mathbf R}_i}\\
+\sum_j \frac{\partial U(\mathbf{R}_i-\mathbf{r}_j)}{\partial {\mathbf R}_i }+d\mathbf{F}_L+\mathbf{\Gamma}_R({\bf R}_i,t),
\end{align}
where $\mathbf{F}_L=\Phi_0 \mathbf{J}/c$ is the Lorentz force. The last term is the Gaussian random force with the correlation function
\begin{equation}\label{eq9}
\langle\Gamma_{R, \mu}({\bf R}_i,t)\Gamma_{R,\nu}({\bf R}_j,t')\rangle=(2 k_B T\eta d)\delta_{i, j}\delta_{\mu, \nu}\delta(t-t').
\end{equation}
where $T$ is temperature, $k_B$ is the Boltzmann constant and $\mu, \nu=x, y$. We assume that  the dynamics of the magnetization is the dissipative one,
\begin{equation}\label{eq10}
\tau\frac{\partial M(\mathbf{r},t)}{\partial t}=-\left[\frac{M(\mathbf{r},t)}{\chi} -B(\mathbf{r})\right]+\Gamma_M(\mathbf{r},t).
\end{equation} 
From Eq.~(\ref{eq10}) we see that relaxation time of the magnetization measured experimentally in the crystal is $\chi\tau$. The last term is the Gaussian random force with 
the correlation function
\begin{equation}\label{eq11}
\langle\Gamma_M({\bf r},t)\Gamma_M({\bf r}',t')\rangle=(2k_B T\tau)\delta({\bf r}-{\bf r}')\delta(t-t').
\end{equation}

\section{Response of the Vortex lattice to an ac driving current}

We consider the response of the vortex lattice to an ac driving current in the flux flow state. In this case, the thermal fluctuations can be neglected. Meanwhile, the quenched disorder can also be neglected because the vortex motion quickly averages out the disorder and the lattice ordering is improved\cite{Koshelev94,Besseling03}. In the absence of thermal fluctuations and quenched disorder, the vortex lattice moves as a whole and the interaction between vortices cancels. The motion of the center of mass for the vortex lattice, $u(t)$ , along the $x$-axis is described by the equation [${\bf r}=(x,y)$]
\begin{equation}\label{eq12}
\eta \frac{du}{dt}=\frac{\partial}{\partial u}\left[\int d{\bf r}B(x+u(t),y,t)M({\bf r},t)\right]+F_L.
\end{equation}
In the Fourier representation of the vortex lattice, we have coupled equations for the center of mass of the vortex lattice $u(t)$ and components of the magnetization $M({\bf G},t)$:
\begin{equation}\label{eq13}
\eta \frac{du}{dt}=n_v\sum_G\left[i G_x \frac{\Phi_0\exp[i G_x u]}{1+\lambda^2 \mathbf{G}^2}M(\mathbf{G},t)\right]+F_L.
\end{equation}
\begin{equation}\label{eq14}
\tau\frac{\partial M(\mathbf{G}, t)}{\partial t}=-\frac{M(\mathbf{G}, t)}{\chi}+\frac{\Phi_0\exp[-i G_x u]}{1+\lambda^2 \mathbf{G}^2},
\end{equation}
where $n_v$ is the vortex density and ${\bf G}$ are the reciprocal vectors of vortex lattice.   The coupling between magnetization and vortices drops as $(1+G^2\lambda^2)^{-2}$ as $G$ increases, thus it is sufficient to take the dominant lattice wave number $\mathbf{G}_1=(2\pi /a, 0)$, where $a$ is the lattice constant\cite{Bulaevskii12a,Lin12m}. Here we consider a dense square lattice and $n_v=1/a^2$, while $G_1\lambda \gg 1$. Renormalizing time in unit of $\tau\chi$, length in unit of $1/G_1$, force per unit vortex length in unit of $\eta /(\tau G_1\chi)$, we have equations for $m(t)=M(G_1,t)\lambda^2G_1^2/(\Phi_0\chi)$ and $u(t)$:
\begin{equation}\label{eq15}
\partial _t m(t)=-[m(t)-\exp [-i u(t)]],
\end{equation}
\begin{equation}\label{eq16}
\partial _t u=F_L-\text{Im}\left[ F_p\exp (i u)m(t)\right],
\end{equation}
with an ac Lorentz force $F_L=F_{\text{ac}}\sin (\omega  t)$ and magnetic pinning force  per unit vortex length $F_p=\Phi _0^2\chi^2\tau/(2\pi^2\lambda ^4\eta )$.  Eliminating $m(t)$ we obtain equation for $u(t)$:
\begin{equation}
\frac{du}{dt}=F_L-F_p\int _0^tdt'\sin[u(t)-u(t')]\exp(t'-t).
\end{equation}
The dc limit $\omega=0$ has been considered in Refs. \onlinecite{Bulaevskii12a,Lin12m}. The behavior of the system depends on $F_p$. For $F_p\geq 8$ the system exhibits hysteretic behavior. Upon increasing the current from zero, at the critical velocity $v_c\approx 1$ the system jumps to the conventional Bardeen-Stephen branch by dissociation of polaron at the threshold 
force $F_L=F_{Lc}\approx 0.5 F_p$. When one reduces the current in the Bardeen-Stephen branch, the system jumps to the branch with larger viscosity by formation of polaron at a retrapping current that is different from the dissociation current. 

\begin{figure}[t]
\psfig{figure=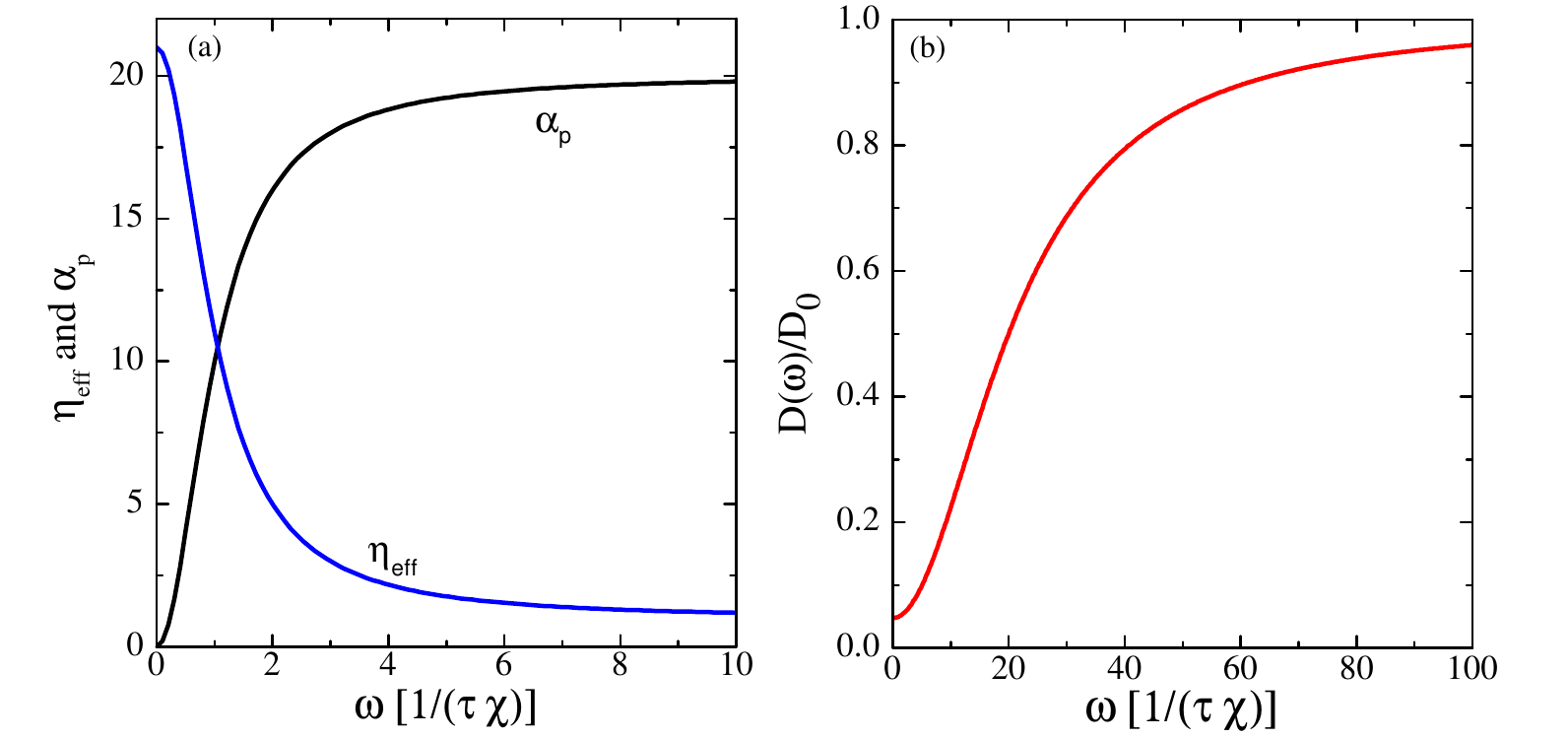,width=\columnwidth}
\caption{\label{f1}(color online) (a) The dependence of effective viscosity $\eta_{{\rm eff}}$, pinning strength $\alpha_p$, and (b) normalized dissipation power, $D/D_0$, on the driving frequency $\omega$ in the linear regime $F_{ac}\ll \omega(1+F_p)$. Here $F_p=20$. }
\end{figure}

In the ac current regime, for a low amplitude $F_{ac}/[\omega(1+F_p)]\ll 1$, the vortex lattice oscillates, $u={\rm{Re}}[u_{ac}\exp(i \omega t)]$, with the amplitude
\begin{eqnarray}\label{eq17}
&&u_{ac}=F_{\text{ac}}(i \eta_{{\rm eff}}\omega +\alpha_p)^{-1}, \\
&&\eta_{{\rm eff}}=1+F_p(\omega^2+1)^{-1}, \ \ \ \ \ \alpha_p=F_p\omega^2(\omega^2+1)^{-1}.
\end{eqnarray}
For a low frequency $\omega\ll 1$, the effect of magnetization is to renormalize the viscosity from $\eta$ to $\eta_{\rm{eff}}=1+F_p$. In this polaron region, the magnetization follows vortex motion by formation of vortex polaron, as in the dc case $\omega=0$, resulting in enhancement of viscosity and suppression of ac dissipation. For a large frequency $\omega\gg 1$, the effect of magnetization is to introduce pinning potential $U_M=F_pu^2/2$ with strength $F_p$. In this case, the vortex lattice follows the driving force much faster than magnetization, which remains almost time independent. 
In the presence of magnetic moments the dissipation power of the whole system, averaged over time, $D(\omega)=\langle F_L(t)v(t)\rangle_t$, is reduced. In the linear region with vortex polaron, we obtain 
\begin{equation}
D=\frac{F_{{\rm ac}}}{2} \frac{\omega^2\eta_{{\rm eff}}}{\alpha_p^2+\eta_{{\rm eff}}^2\omega^2}.
\end{equation}
This dissipation power should be compared with that, $D_0F_{{\rm ac}}^2/2$,  without magnetic moments (at $F_p=0$). For $\omega\ll1$ we get $D/D_0=(1+F_P)^{-1}$ and for $\omega\gg 1$
we get $D/D_0\approx 1$. The frequency dependence of the effective viscosity, $\eta_{{\rm eff}}$, pinning strength $\alpha_p$ and normalized dissipation power $D/D_0$ is shown in Fig.~\ref{f1}.

The polarization of the magnetization results in periodic pinning potential with the periodicity of  vortex lattice as it was induced by the same lattice at previous positions and previous moments of time.  
The important point for possible applications is that in the presence of magnetic subsystem, the dissipation of the system is strongly reduced in linear regime $F_{{\rm ac}}<F_{Lc}$. 

\begin{figure}[b]
\psfig{figure=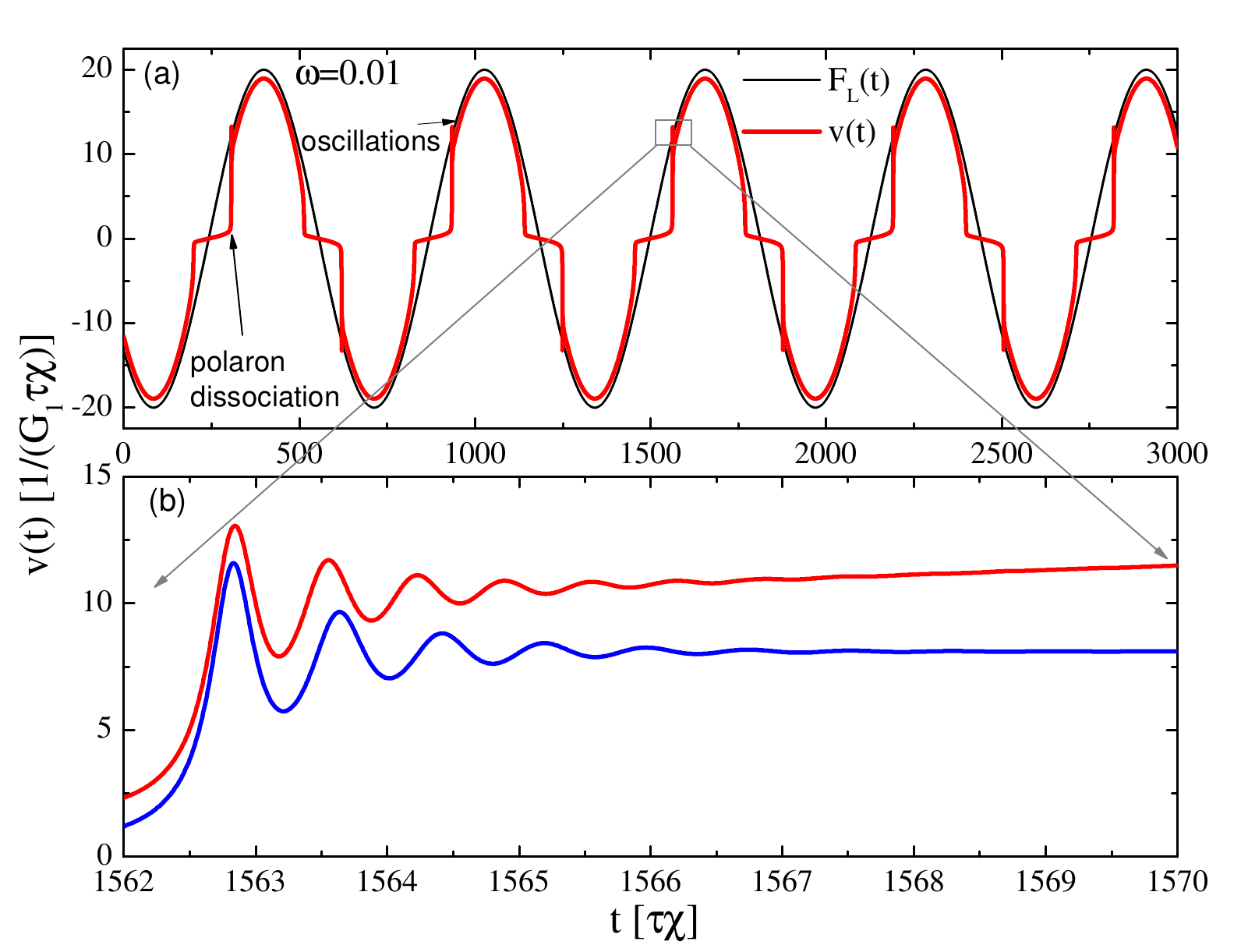,width=\columnwidth}
\caption{\label{f2}(color online) (a) $v(t)$ and $F_L(t)$ as a function of time  obtained by numerical solution of Eqs. (\ref{eq15}) and (\ref{eq16}) with $\omega=0.01$, $F_{{\rm ac}}=20$ and $F_p=20$.(b) Red line:close up for the fast oscillations of velocity in (a). Blue line numerical solution of Eq.~ (\ref{eq177}) with $F_{Lc}=8.103$. }
\end{figure}

For a stronger driven force amplitude, $F_{{\rm ac}}>F_{Lc}$, in the hysteretic regime, we describe the system analytically  in the adiabatic limit, $\omega\ll 1$. At the time 
moment $t_c$, when $F_L(t)=F_{Lc}\approx  0.5 F_p$, polaron dissociation leaves magnetization and vortex lattice weakly coupled because lattice moves now with a high velocity. The magnetization  component 
$m(t)$ after that moment  relaxes as $m(t)=\exp(-t+t_c)$, and motion of vortex lattice is determined by the equation
\begin{equation}\label{eq177}
\frac{du}{dt}=F_{Lc}+F_p\sin ( u)\exp(-t+t_c).
\end{equation}
When $t-t_c> 1$ the velocity of the vortex lattice oscillates with the frequency $\Omega=F_{Lc}$ but oscillations relax on the time scale of unity, 
\begin{equation}\label{eq178}
v=F_{Lc}+F_p\sin(F_{Lc} t)\exp(-t+t_c).
\end{equation}
These post dissociation oscillations are due to the motion of vortex lattice in periodic potential induced by remnant retarded magnetization. Results for arbitrary time $t-t_c$ are shown in Fig.\ref{f2}, which are compared with the direct numerical solution of Eqs. (\ref{eq15}) and (\ref{eq16}).

\begin{figure}[t]
\psfig{figure=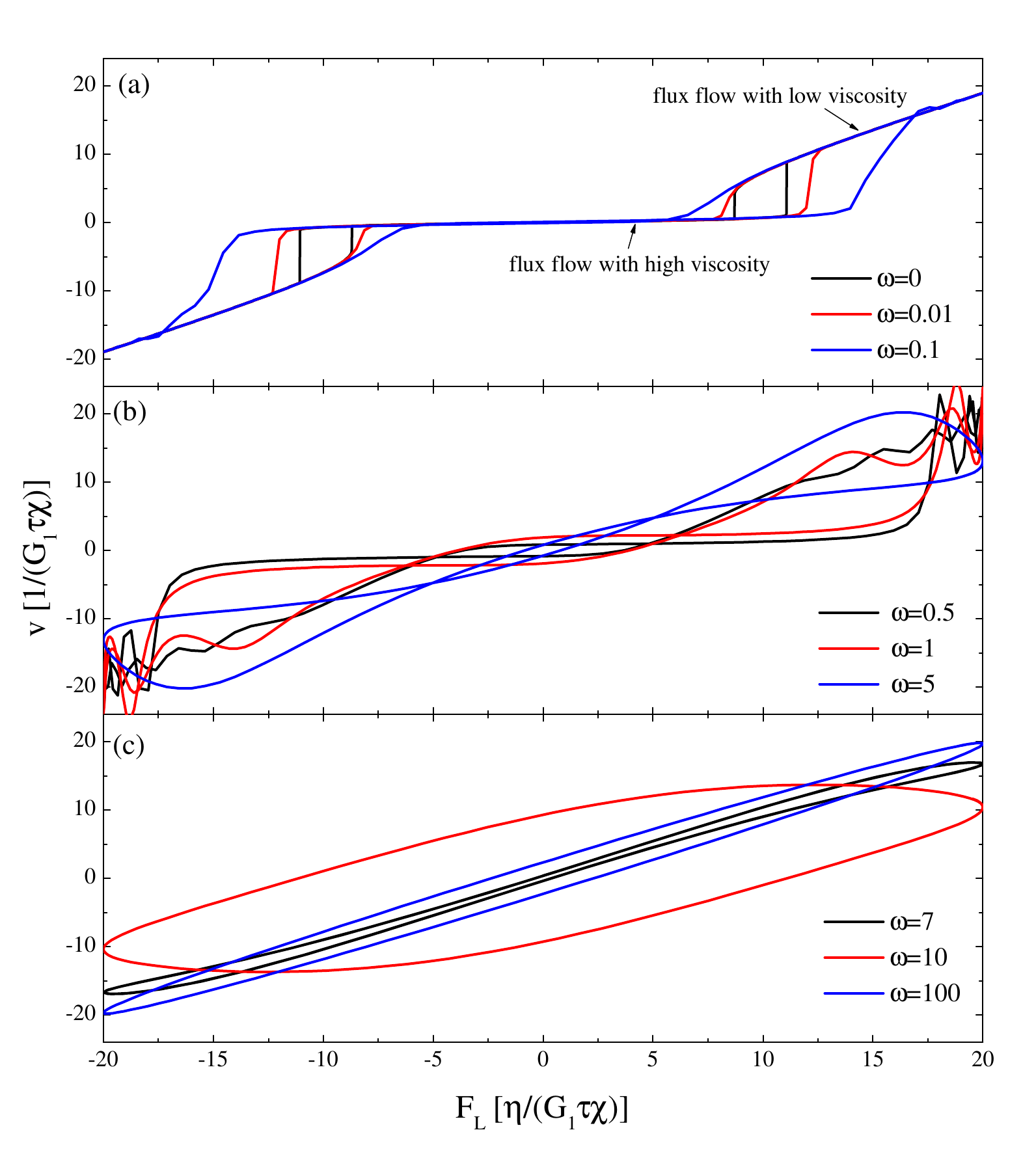,width=\columnwidth}
\caption{\label{f3}(color online) Dependence of the vortex velocity $v(t)$ on the driving force $F_L(t)=F_{{\rm ac}}\sin(\omega t)$ with $F_{{\rm ac}}=20$. Here $F_p=20$.}
\end{figure}

\begin{figure}[b]
\psfig{figure=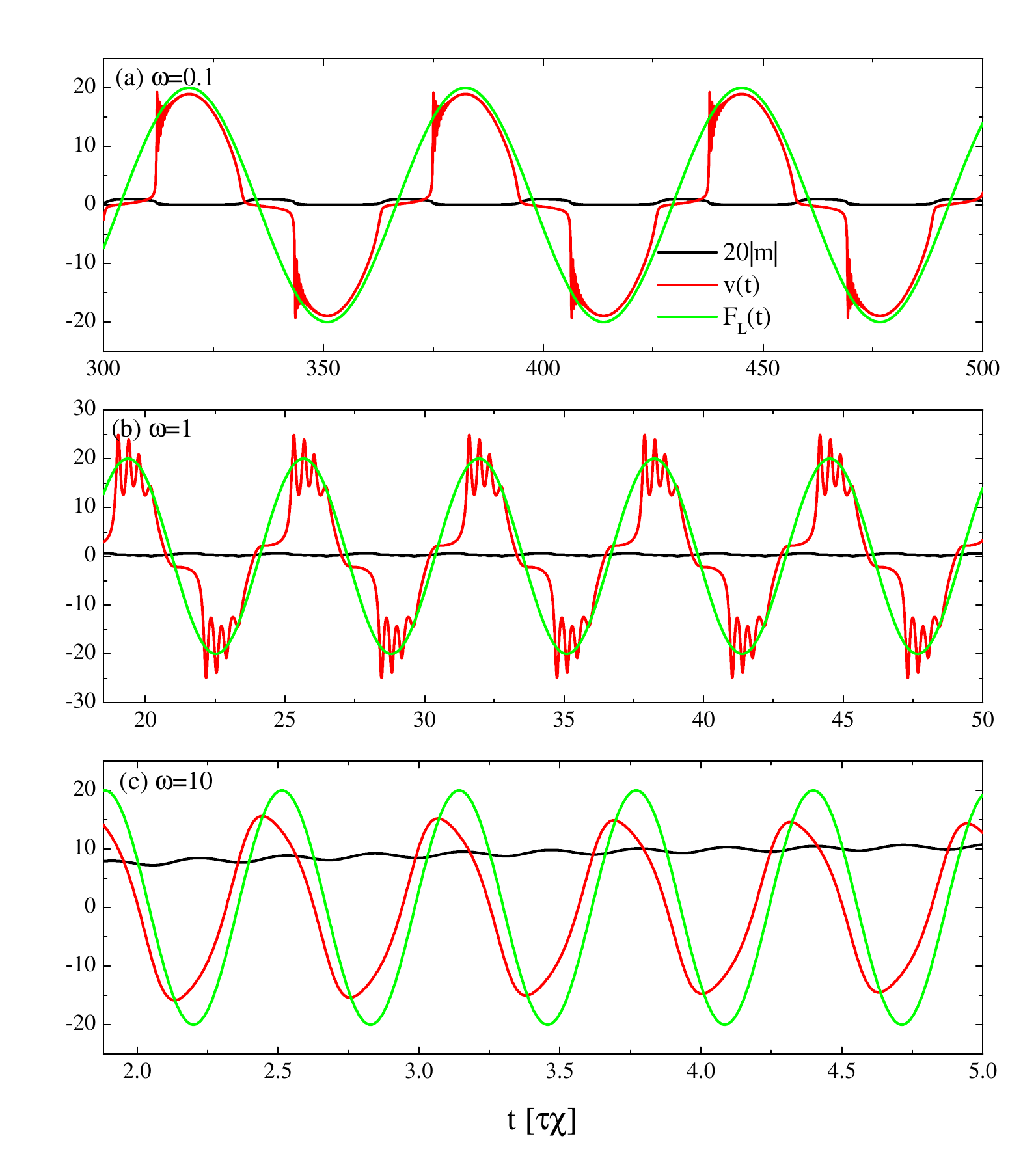,width=\columnwidth}
\caption{\label{f4}(color online) Time evolution of the vortex velocity, $v(t)$, and of magnetization $|m(t)|$ in the presence of ac driving force $F_L(t)=F_{{\rm ac}}\sin(\omega t)$ at several frequencies $\omega$. (a): $\omega=0.1$, (b): $\omega=1$, (c): $\omega=10$. We take $F_{{\rm ac}}=20$ and $F_p=20$.}
\end{figure}

For arbitrary $\omega$
we solve numerically Eqs. (\ref{eq15}) and (\ref{eq16}) taking both the retardation and nonlinearity into account. We consider the interesting region $F_p\geq 8$, where the dissociation of vortex polaron is possible due to nonlinear effects at $u\geq 1$. We take $F_p=20$ in the following discussion. The hysteretic behavior of vortex lattice velocity vs. driving force is shown in Fig.~\ref{f3}. At frequencies $\omega\lesssim 1$,  during the period of $F_L(t)$, we see the following sequence of events: polaron formation near low $|F_L|$ (interval of low vortex velocity), then polaron dissociation (velocity sharp increase) followed by region of vortex oscillations on the background of average high velocity, decrease of velocity as the Lorentz force drops and vortex retrapping (sharp drop in vortex velocity) and then dissociation again at negative $-F_{Lc}$ (sharp drop in velocity). The results for 
behavior of vortex velocity in time, $v(t)$, at $F_{ac}=10>F_{Lc}$ and different $\omega$ are shown in Fig. ~\ref{f4}. 

\begin{figure}[t]
\psfig{figure=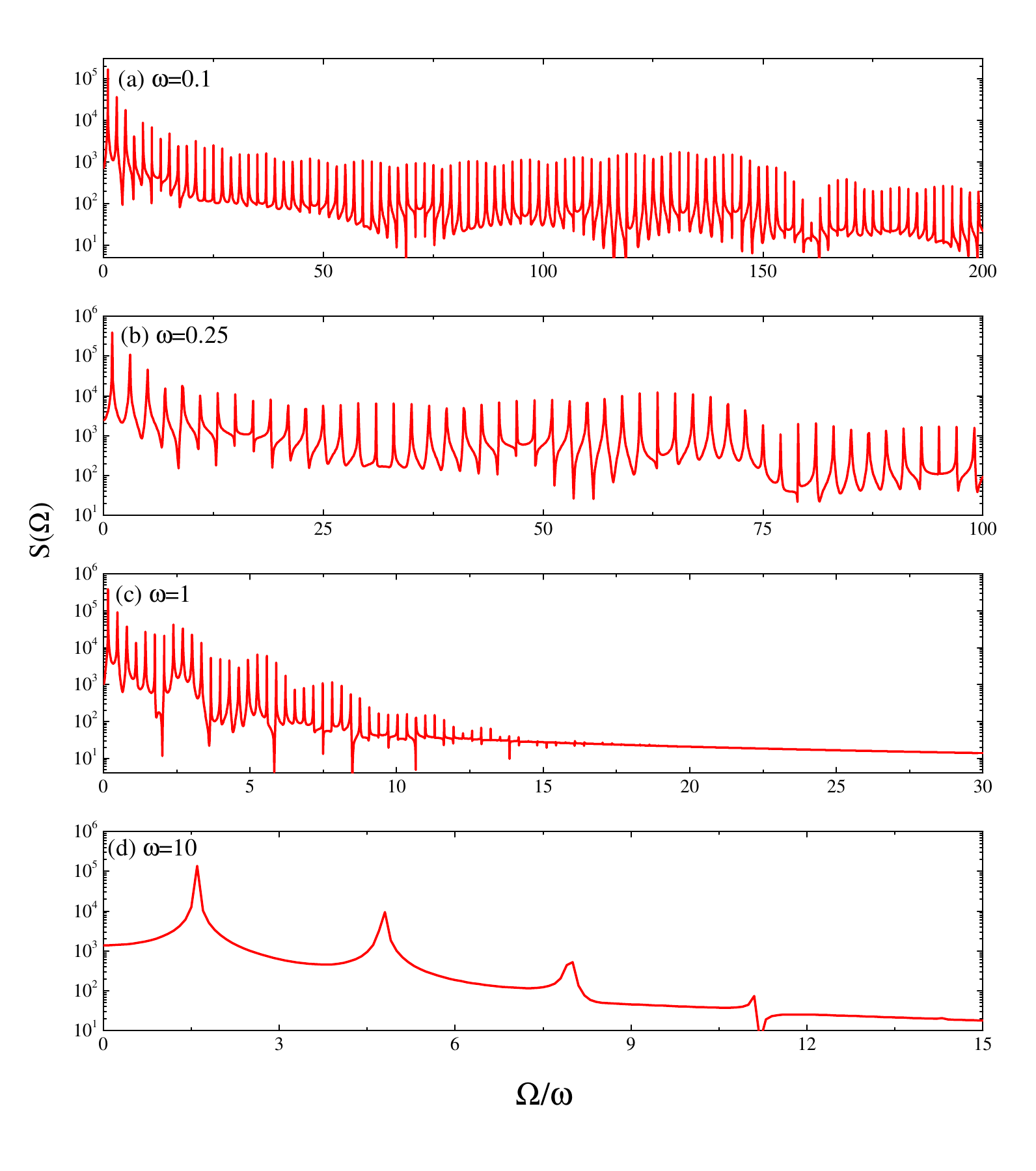,width=\columnwidth}
\caption{\label{f5}(color online) Normalized spectral function $S(\Omega)$ of vortex lattice oscillations at different frequencies of the driving force $\omega =0.1, 0.25, 1, 10$ at the amplitude $F_{{\rm ac}}=20$ and $F_p=20$. }
\end{figure}

At all frequencies $\omega\lesssim 1$ we see post dissociation oscillations caused by motion of decoupled  vortices with respect to periodic potential created by nonuniform magnetization  
induced by the same lattice just before decoupling (when velocity was still low) and frozen for some periods of time after decoupling due to the retardation effect. This self-induced pinning, resulting from the retardation, and the amplitude of corresponding vortex oscillations reach maximum at $\omega\approx 1$. In a rough approximation we describe them by the equation
\begin{equation}
\frac{du}{dt}\approx F_{Lc}+F_p m_d\sin(u-u_d),
\end{equation} 
assuming approximately constant $m$ and $F_L=F_{Lc}$ in the regions of maxima and minima of the Lorentz force. Here $u_d$ and $m_d$ are the position of vortex lattice and the amplitude of magnetization at the moment of decoupling.
This gives approximate solution near extrema of $F_L(t)$ at $F_{{\rm ac}}$:
\begin{equation} 
v(t)\approx F_{Lc}+F_pm_d\sin(F_{Lc}t),
\end{equation}
which provides rough estimate for the oscillation frequency, $\Omega\approx F_{Lc}$,  when number of oscillations of velocity per the half period of $F_L(t)$ is significantly bigger than unity. This expression for the frequency in original units reads $\Omega\approx 2\pi F_{{\rm ac}}/a\eta$. Such relation is anticipated for decoupled vortex moving in the pinning potential with periodicity $a$. In Fig.~\ref{f5} we show normalized spectral function  $S(\Omega)=|v^2(\Omega)|/F^2_{{\rm ac}}$ for different driving force frequency at $F_{{\rm ac}}=20$ [we denote $v(\Omega)=\int dtv(t)\exp(i\Omega t)$]. The spectrum of oscillations consists of harmonics  $(2n+1)\omega$ due to nonlinearity of the equation for $u$ ($n$ is integer) and amplitudes of these harmonics are enhanced near the frequency   $\Omega\approx F_{Lc}$ when $\omega\lesssim 1$.

\begin{figure*}[t]
\psfig{figure=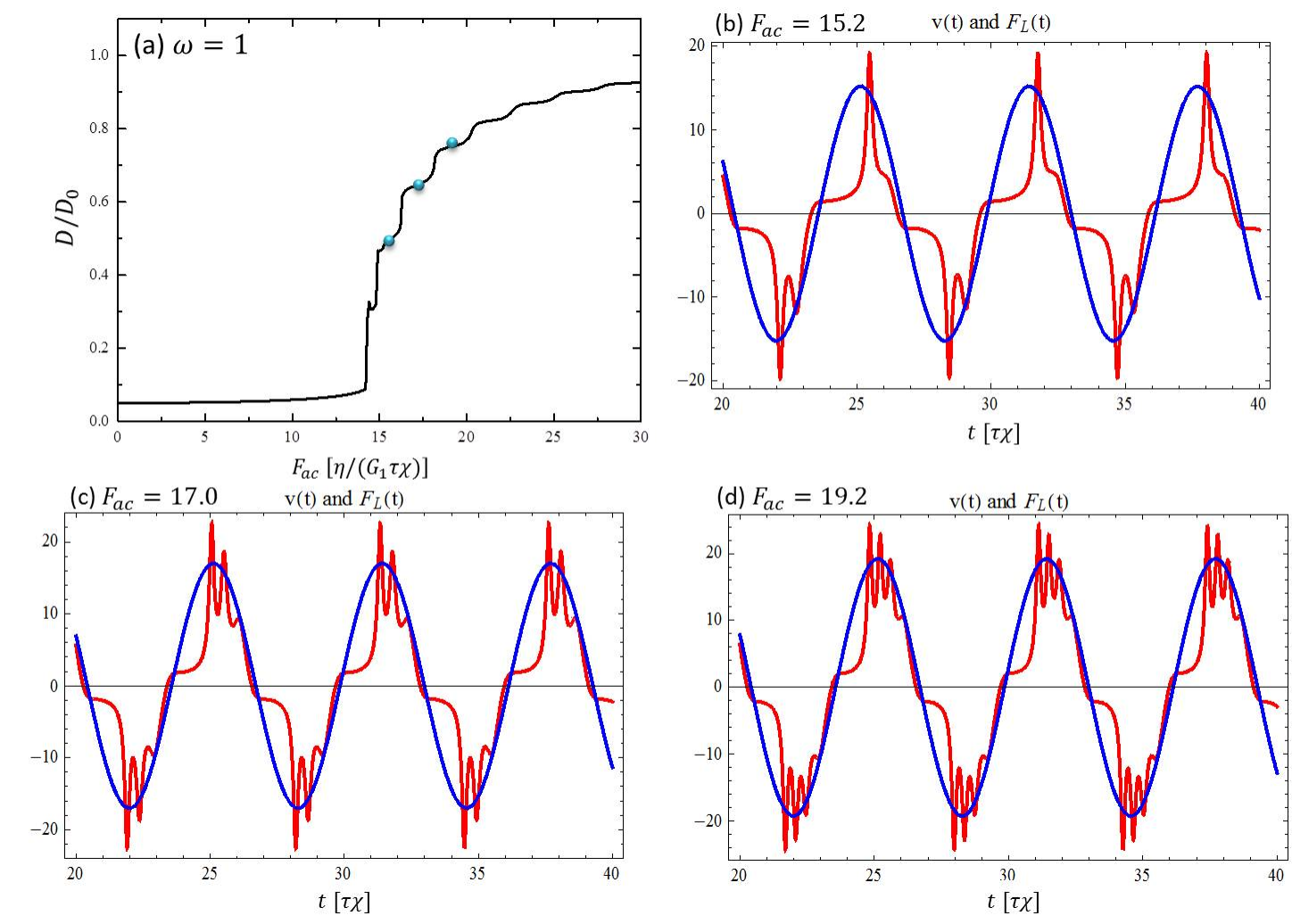,width=18cm}
\caption{\label{f6}(color online) The normalized dissipation $D(F_{{\rm ac}})$ as a function of the amplitude of the driving force and time evolution of $v(t)$ and $F_L(t)$,  at several values of $F_{{\rm ac}}$, corresponding to the different plateaus in the dissipation power $D/D_0$ in (a). }
\end{figure*}
\begin{figure}[t]
\psfig{figure=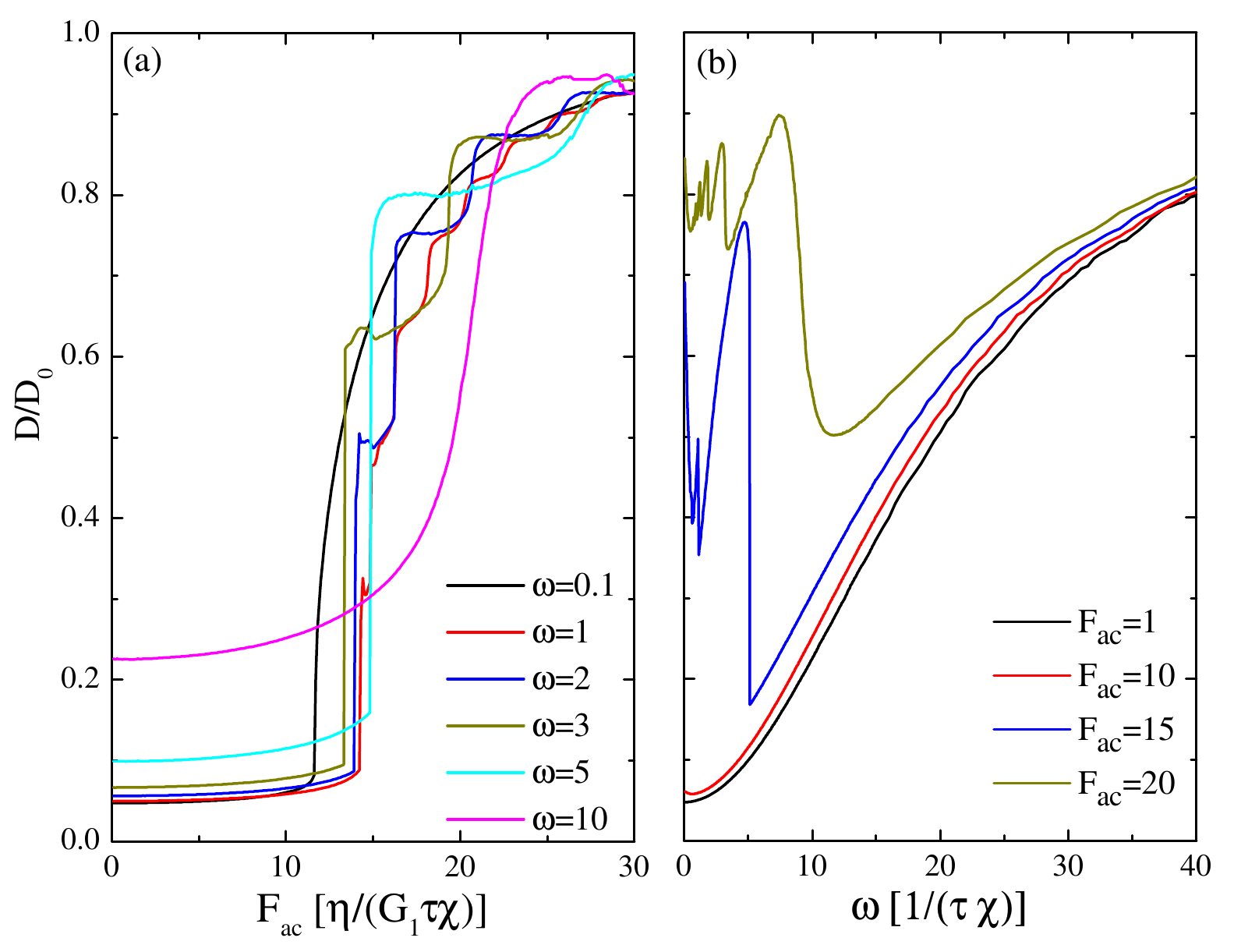,width=\columnwidth}
\caption{\label{f7}(color online) Dependence of the normalized dissipation power $D/D_0$ on 
$F_{{\rm ac}}$ in (a) and on $\omega$ in (b). Here $F_p=20$.}
\end{figure}

In the oscillation regime part of dissipation power, $\langle v(t)F_L(t)\rangle$, is transferred into oscillation power in the frequency region near $\Omega=F_{Lc}$. This part increases with the number of oscillations per half of the Lorentz force period. In Figs.~\ref{f6} we show dependence of normalized dissipation power, $D(F_{{\rm ac}})/D_0$, as a function of $F_{{\rm ac}}$
and the corresponding behavior of $v(t)$ and  $F_L(t)$. For $\omega<1$, dissipation increases step wise as $F_{{\rm ac}}$ increases. To understand the plateaus and steps in dissipation power, we plot $v(t)$ and $F_L(t)$ at several $F_{{\rm ac}}$ corresponding to plateaus in $D/D_0$. As shown in Fig.~\ref{f6}, the different plateaus correspond to different number of periods of the fast oscillations during the time $\pi/\omega$. Hence, these plateaus are due to the commensurability in periods of fast oscillations and slow oscillations of the driving force. The sharp increase of dissipation power from one plateau to another is caused by changing of the period of fast oscillation.  The behaviour in the region $F_{ac}/\omega\gg 1$ is highly nonlinear, and numerical calculations become important tool to understand the dissipation in the system.

For high frequency $\omega\gg 1$ it is impossible to accommodate fast oscillation with frequency $F_{{\rm ac}}$ during the period of the external driving frequency, and step structure disappears, as shown in Fig. \ref{f7}. For high frequency $\omega$ the magnetization $m(t)$ is approximately periodic in space (as induced by periodic vortex lattice) and oscillates in time  with the amplitude which decreases with $\omega$ as shown in Fig.~\ref{f4}.  
 
Hence, at low driving force amplitudes, $F_{{\rm ac}}<0.5 F_p$, the polaronic effect renormalizes the viscosity at low $\omega$ and introduces pinning at high $\omega$. At a long magnetic relaxation time, $\omega\tau\chi\gg 1$,
the elastic constant of the magnetic pinning force per unit vortex length in dimension form becomes $\alpha_p=F_p \eta /(G_1\chi\tau)=\Phi_0^2\chi/(2\pi^2\lambda^4G_1)$, while at small $\omega\tau\chi \ll 1$
the vortex viscosity is enhanced from $\eta$ to $\eta(1+ F_p)$. At higher values of $F_{{\rm ac}}$ reduction of dissipation is lower. In Fig.~\ref{f7} we show the normalized dissipation power as a function of $F_{{\rm ac}}$ and $\omega$. Due to commensurability effects as shown in Fig. \ref{f6}, $D_0/D$ oscillates with $\omega$ in the region $\omega\sim 1$ when $F_{ac}>F_{Lc}$ as shown in Fig. \ref{f7} (b). For a large $\omega$ or $F_{ac}<F_{Lc}$, $D_0/D$ increases smoothly with $\omega$.

To understand significance of polaron pinning we compare the elastic constant of pinning potential for vortex $\alpha_p=F_p \eta/(\tau\chi)$ with the maximum possible $\alpha_p$ for a single vortex pinning.  Such a maximum, $\alpha_{c}\approx H_c^2/(8\pi)=\Phi_0^2/(64\pi^3\lambda^2\xi^2)$ is 
reached in the case of columnar defects produced by irradiation of crystal by heavy ions\cite{Civale91}, where we have assumed that the size of defects is $\xi$, while $H_c$ the thermodynamic critical field. 

For ErNi$_2$B$_2$C at temperature 3 K, $\lambda=500\ \AA$, $\xi=135\ \AA$, $\chi\approx 0.05$, $\rho_n=10^{-8}\ \rm{\Omega\cdot m}$ and we estimate $\alpha_p/\alpha_c\approx 0.4$. \cite{Yaron1996,Cava1994, Bonville1996}  
Thus magnetic pinning is not drastically lower than maximum possible one.

To observe hysteretic behavior and high frequency oscillations of voltage following the dissociation, one needs to exceed the critical current $J_c$ corresponding to $F_{Lc}$. In $\rm{ErNi_2B_2C}$ this critical current at $\omega\rightarrow  0$ is quite high, about $10^6$ A/cm$^2$ at low temperatures for the field 0.1 T along the a or  b crystal axis. It increases with $\omega$ as shown in Fig.~\ref{f3}. However, $J_c$ proportional to $\sin^2\alpha$ becomes smaller , as field approaches the $c$-axis, due to renormalization of $\chi$ as $\chi\sin^2\alpha$. Here $\alpha$
is the angle between the applied magnetic field and the $c$ axis. 

The important parameter which determines the dynamics of polaron vortices in magnetic superconductors is 
\begin{equation}
F_p=\frac{\tau\chi^2\xi^2 c^2\rho_n}{\pi\lambda^4}.
\end{equation}
It depends crucially on the magnetization relaxation time $\chi\tau$, measured experimentally, and the magnetic susceptibility $\chi$ which are only known for the most studied Er borocarbide. In this crystal according to M\"{o}ssbauer measurements\cite{Bonville1996} $\chi\tau\approx 5\times 10^{-10}$ s at $T=5$ K and we anticipate longer time at lower temperatures.  For $\chi=5\times10^{-2}$ and $\rho_n=5$ $\mu\Omega\cdot$cm we get $F_p> 120$. In Tm borocarbide $\rm{TmNi_2B_2C}$ we anticipate bigger $\chi$ at temperatures above Neel temperature about 1 K due to larger density of almost free spins, but $\tau$ for Tm borocarbide is unknown.

As a consequence of polaron dissociation at driving force with high amplitude and not very high frequency vortex lattice velocity oscillates and generates ac voltage with frequency of the driving force and also high frequency voltage whose frequency depends on the driving force amplitude. These high frequencies are about 10 - 20 times bigger than driving force frequency and conversion of the power to these frequencies is quite efficient at $\omega<1$ and $F_{{\rm ac}}\approx 20$. 
This implies that the vortex motion in magnetic superconductors may be used to generate high frequency electromagnetic oscillations whose frequency depends on the amplitude of the driving current.

\section{Vortex creep}

\begin{figure}[t]
\psfig{figure=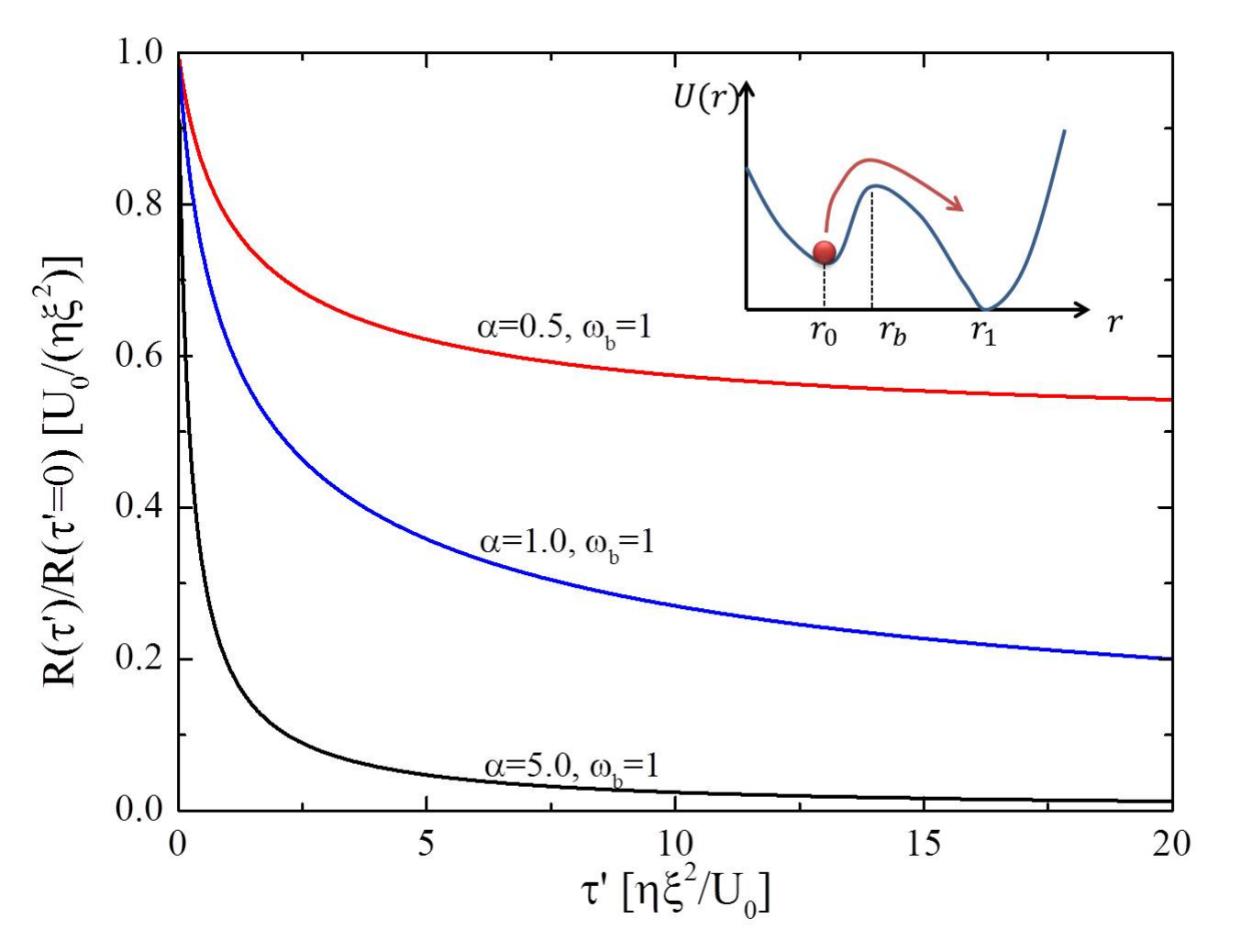,width=\columnwidth}
\caption{\label{f8}(color online) Renormalized rate $R(\tau')/R(\tau'=0)$ as function of $\tau'$ at several $\alpha$. Here we take $\omega_b=1$. Inset is a schematic view of thermally activated vortex escaping from a pinning potential.}
\end{figure}

Next we discuss how polarization of the magnetic moments affects the vortex creep. We consider motion of a single vortex from one pinning center at $r_0$ to another pinning center at $r_1$ due to the thermal fluctuations, as schematically shown in the inset of Fig. \ref{f8}. We assume that $U(r_0)>U(r_1)$. The behavior of vortex is characterized by a rate for thermal escape of vortex from a metastable potential. In the absence of the magnetization, the rate can be calculate using the standard Kramers' theory\cite{Hanggi1990}. The magnetization with relaxation dynamics introduce memory effect for the vortex crossing the barrier. Here, we investigate the effect of retardation on the escape rate. 

Without a Lorentz force, the equation of motion for vortex is
\begin{equation}\label{eq188}
\eta \partial _t r=-\partial _rU+\int d{\bf k} \frac{i k_x M(k)}{2\pi}\frac{\Phi _0 \exp (i k_x r)}{\lambda ^2{\bf k}^2+1} +\Gamma_R,
\end{equation}
\begin{equation}\label{eq199}
\tau \partial _tM(k)=-\left(\frac{M(k)}{\chi}- \frac{\Phi _0}{2\pi}\frac{\exp [-i k_x r]}{\lambda ^2\mathbf{k}^2+1}\right)+\Gamma_M.
\end{equation}
Main contribution to integral in Eq. (\ref{eq188}) comes from the region $k<1/\lambda$. We assume that the distance between pinning potentials is much smaller than $\lambda$, i.e. $|r_1-r_0|\ll \lambda$. Then we can expand in $k_x r$ in Eqs. (\ref{eq188}) and (\ref{eq199}). We then obtain the following equation of motion for vortex
\begin{eqnarray}
&& \partial _tr=-\partial _rU+\alpha(M_1- r)+\Gamma_r, \label{eq26}\\
&&\chi \tau \partial _tM_1=-\left(M_1-  r\right)+\Gamma_1, \label{eq27}
\end{eqnarray}
with the correlation functions
\begin{eqnarray}
&&\left\langle\Gamma_r(t)\Gamma_r(t')\right\rangle=2 T\delta(t-t'), \label{eq28}\\
&&\left\langle\Gamma_1(t)\Gamma_1(t')\right\rangle=2 T\tau \delta(t-t')/\alpha, \label{eq29}
\end{eqnarray}
where we have introduced the following dimensionless units: $r$ and $M_1$ is in unit of $\xi$, $U$ is in unit of $U_0$, time is in unit of $\eta  \xi ^2/U_0$, temperature is in unit of $U_0 d/k_B$. 
The strength of coupling between magnetization and vortex is $\alpha=A\xi^2/U_0$. 
Here 
\begin{eqnarray}
&&M_1=\frac{\Phi_0}{2\pi A}\int d{\bf k}\frac{ik_xM({\bf k})}{\lambda^2{\bf k}^2+1}, \\
&&A=\int d{\bf k}\frac{\chi\Phi_0^2}{(2\pi)^2} \frac{k_x^2}{\lambda^2{\bf k}^2+1}\approx\frac{\chi\Phi_0^2}{4\pi\lambda^4}\left[\ln\left(\frac{2\pi\lambda}{\xi}\right)-0.5\right].
\end{eqnarray}
From Eq. (27), we obtain for $M_1$
\begin{equation}\label{eq30}
M_1(t)=r(t)-e^{-t/\tau '}\int _0^te^{t'/\tau '}\left[\partial _tr(t')+\frac{\Gamma _1(t')}{\tau '}\right]dt',
\end{equation}
with $\tau'=\chi\tau$. Substituting Eq. (\ref{eq30}) into Eq. (\ref{eq26}), we obtain
\begin{equation}\label{eq31}
\int _0^tK(t-t')\partial _tr(t')dt'= -\partial _rU+\Gamma ',
\end{equation}
with
\[
K(t-t')=\alpha   \exp [-|t-t'|/\tau ']+\delta (t-t'),
\]
\[
\Gamma ' =\alpha   e^{-t/\tau '}\int _0^te^{t'/\tau '}\frac{\Gamma _1(t')}{\tau '}dt'+\Gamma _r.
\]
Note that the nonlocal dissipation and noise still satisfy fluctuation-dissipation theorem
\[
\left\langle\Gamma'(t)\Gamma'(t')\right\rangle=2 T K(t-t').
\] 
Escaping rate for viscosity with retardation effect was calculated by H\"{a}nggi and Mojitabai in Ref.~\onlinecite{Hanggi1982}. We use their results for our case of the viscosity containing parts with and without retardation. We obtain the expression
\begin{equation}\label{eq32}
R(\tau')=\frac{\omega _0\beta }{2\pi  \omega _b}\exp [-\Delta U/T],
\end{equation}
\begin{equation}\label{eq33}
\beta =\frac{-1-\alpha  \tau '  +\tau ' \omega _b^2+\sqrt{4 \tau ' \omega _b^2+\left(1+\alpha  \tau '  -\tau ' \omega _b^2\right){}^2}}{2 \tau '},
\end{equation}
where $\omega_0=\sqrt{\partial_r^2 U(r_0)}$, $\omega_b=\sqrt{-\partial_r^2 U(r_b)}$, and $\Delta U=U(r_b)-U(r_0)$. Here $r_b$ is the position of the barrier top. The results for $R(\tau')/R(\tau'=0)$ are shown in Fig. \ref{f8}. The rate decreases with $\tau'$.
Equation (\ref{eq32}) is valid if $R(\tau')\tau'\ll 1$. In the limit of strong interaction of vortex with magnetic system, $\alpha>\omega_b^2 $, this expression is valid at any $\tau$ because $\beta$ drops with $\tau$ as $1/\tau^2$. In this case the barrier remains the same as in the system without magnetic moments (at $\alpha=0$), but pre-exponential factor diminishes as $\tau$ increases due to interaction of vortex with magnetic moments. 

For weak interaction, $\alpha<\omega_b^2 $, the rate has nonzero limit at a large $\tau$, and the condition $R(\tau')\tau'\ll 1$ may 
be violated at a large $\tau$. When it occurs, the barrier is renormalized from $\Delta U$ to $\Delta U+\alpha(r_b-r_0)^2/2$ as magnetization cannot follow escaping vortex. Pre-exponential remains the same as the bare vortex system with $\alpha=0$. 

In both cases the rate decreases due to the polaronic effect. Effect depends strongly on the value $\alpha$ which is unknown for $\rm{ErNi_2B_2C}$ at this moment because the characteristics of the pinning centers are unknown. Measurements of the relaxation of magnetization $d\ln M/d\ln t$ above 6 K, where the polaron mechanism is absent, may reveal the pinning barriers in the crystals. Then one can predict suppression of the creep rate at lower temperatures. 

\section{Conclusion}

In conclusion, the polaronic effect in magnetic superconductors (crystals and multilayered structures) suppresses dissipation for ac currents. This effect depends strongly on the parameter $\omega\tau\chi$. When this parameter is large, magnetic moments result in additional strong pinning due to the polaronic effect. Creep rate at low temperatures is reduced as well. Thus magnetic superconductors may be useful for applications, especially for devices with ac currents. 
We predict the generation of voltage oscillations with high frequency $\Omega\gg\omega$ in the presence of strong ac driving Lorentz force with moderate frequency $\omega\sim 1/(\tau\chi)$.  This effect is caused by pinning of vortex lattice on retarded periodic magnetization induced by the same vortex
lattice in previous moments of time.

\acknowledgments
The authors would like to thank  C. Batista, V. Vinokur and A. Saxena  for helpful discussions.
This publication was made possible by funding from the Los Alamos Laboratory Directed Research and Development Program, project number 20110138ER.

%

\end{document}